**Title: Trapping Light in Plain Sight: Embedded Photonic Eigenstates in Zero-Index Metamaterials**

*Francesco Monticone[1,2], Hugo M. Doeleman[3], Wouter Den Hollander[3], Femius Koenderink[3], and Andrea Alù[1,*]*

*Corresponding Author: E-mail: alu@mail.utexas.edu

[1]The University of Texas at Austin, Department of Electrical and Computer Engineering, Austin, Texas, 78712, USA.
[2]Cornell University, School of Electrical and Computer Engineering, Ithaca, NY, 14850, USA.
[3]FOM Institute AMOLF, Amsterdam, The Netherlands

Confining electromagnetic energy is crucial to enhance light-matter interactions, with important implications for science and technology. Here, we discuss the opportunities offered by trapping and confining light in open structures, based on the concept of embedded eigenstates within the radiation continuum enabled by zero-index metamaterials. Building upon the physical insights offered by our analysis, we put forward a general platform that allows realizing extremely high field enhancements in open structures under external illumination. Structures supporting embedded eigenstates represent a rare example of physical systems in which extreme – in principle unbounded – responses can be tamed. Our proposed design recipe to realize bound states in the continuum also offers a simple model that allows testing important questions that surround the concept of embedded eigenstates, such as their effect on the local density of photonic states. Our findings help clarifying which nano-optical and radio-wave applications may benefit from this unusual and singular response.

**1. Introduction**

As one of the tenets of classical physics, when a point charge is accelerated it radiates electromagnetic energy as governed by Maxwell's equations [1]. This basic principle is at the foundation of a large variety of phenomena and technological advances. Interestingly, however, several *extended* charge distributions have been identified, which can be accelerated without producing any radiation, seemingly at odds with our physical intuition. For example, Ehrenfest pointed out that, due to symmetry considerations, a translationally symmetric sheet of charges or a rotationally symmetric distribution of charges do not radiate when they oscillate in a purely transverse or radial fashion, respectively [2]-[3]. Interest in non-radiating sources has resurfaced several times in the past century, from the foundation of atomic theory, to inverse scattering problems, and recent investigations in metamaterials [4]-[12]. A related but independent research topic in optics and acoustics is the one of *embedded eigenstates*, which are defined as optical states supported by open cavities that are bound and ideally confined (their radiative lifetime and Q factor diverge), despite the presence of available and symmetry-compatible radiation modes (i.e., they exist within the radiation continuum) [13]-[25]. In other words, embedded eigenstates are non-radiating *eigenmodes* supported by open cavities, in analogy to anomalous electronic bound states that can exist in certain quantum systems [26]-[27].

Different from classical investigations on non-radiating sources, which typically consider *impressed* sources, research on embedded eigenstates focuses on distribution of sources that can be *induced* in a material body by an external excitation. This fact adds further complexity to the problem, since we are not free to design charge and current distributions at will, but we are constrained by the response of a material body, and by the passivity and causality bounds of the scattering process. Because of reciprocity in the source-field relation, no ideally-non-radiating eigenmodal current distribution can be excited from a far-field excitation; however, in the case of embedded eigenstates, there is interestingly no bound on how close we can approach the non-radiation condition from far-field [13]-[25]. Remarkably, this means that, in the limit of vanishing material losses, we can design a scatterer that, when externally illuminated, supports an optical state with diverging lifetime and infinite stored energy, which is reflected in the scattering spectrum as a resonance with diverging Q factor and infinitesimally small bandwidth. The realization of such embedded eigenstates represents an intriguing mechanism for extreme light confinement and trapping, drastically different compared to conventional ways to confine light by mirrors, total internal reflections, etc., with important practical implications. A few platforms have been recently identified to realize embedded eigenstates with different dimensionalities, including 2D photonic crystal slabs [13]-[17], 1D arrays of spheres or rods [19]-[20], 3D layered plasmonic nanospheres [21]-[22], among others [25]. Such different implementations show that embedded eigenstates are a general, yet anomalous, feature of open wave-guiding systems – systems that usually exhibit a complex eigenmode spectrum due to energy leakage in the form of radiation [24]. The disappearance of leakage in non-symmetry-protected embedded eigenstates is typically understood, at least mathematically, as a result of ideal destructive interference between radiation from different modes or resonators [25]. However, a solid, bottom-up design strategy that can be applied to systematically realize embedded eigenstates of any dimensionality is yet to be found. To solve this problem, in this paper, we start by presenting general considerations on the origin of embedded eigenstates in different geometries. We show that wave propagation in *periodic* structures, such as the photonic crystal slabs usually considered in the literature, is not a necessary condition to realize these anomalous bound states, which can also be implemented in transversely homogenous (or homogenizable) designs under certain conditions. In light of these findings, we put forward a platform based on zero-index materials and metamaterials that allows realizing embedded eigenstates in different settings, and we propose a route to practically implement these structures at microwave frequencies using near-lossless zero-index metamaterials. In addition, our proposed design provides a simple model that allows testing important questions about the physics and usefulness of embedded eigenstates. For example, we calculate the local density of state of a structure supporting bound states in the continuum, and highlight its limited impact in nanophotonic applications for planar configurations. Our results shed new light on the intriguing phenomenon of embedded eigenstates in electromagnetics, and allow assessing the impact of these concepts for different practical applications, such as energy harvesting, light trapping, control of nano-source radiation, enhanced absorption, nanolasing and sensing.

**2. General Considerations on Embedded Eigenstates**

In this section, we aim at deriving some general principles of embedded eigenstates from the different examples reported in the recent literature on this topic. These principles will guide our design strategy to realize embedded eigenstates in different settings based on zero-index metamaterials. Consider the structures sketched in **Figure 1**, which all differ due to symmetry and dimensionality, yet they can all support bound states within the radiation continuum. How does their geometry affect the onset and nature of these eigenstates? All these systems are assumed to be electromagnetically open, namely, an optical state supported by these structures is in principle allowed to couple with far-field radiation, and vice versa. In the case of 3D open structures (**Figure 1a**), it can be shown [21]-[22] that trapped states within the radiation continuum (i.e., with infinite radiative Q factor) can only be supported if the geometry diverges or collapses (e.g., a dielectric open cavity of infinite radius, or a lossless plasmonic cavity of infinitesimal size), or, in the case of finite bounded material bodies, if the structure involves materials whose permittivity and/or permeability goes to zero or infinity at a given frequency, as shown in [21]. In this setting, an embedded eigenstate corresponds to an eigenmode of the system with real eigenfrequency, i.e., no radiation damping, which is possible only if the fields outside the object are identically zero, as spherical harmonics are always radiative [10],[21]-[23].

In 1D or 2D open structures with continuous or discrete translational symmetry, as in the examples shown in **Figure 1b-c**, light confinement is typically attained as a result of transverse momentum conservation: a mode propagating along an open waveguiding structure (e.g., a dielectric slab, or an optical fiber) with phase velocity $v_p = \omega/\beta$ smaller than the speed of light in the surrounding medium $c_0$ (slow wave) cannot match the transverse momentum of any radiation mode, and therefore the mode remains confined in the waveguide (here, $\omega$ indicates the angular frequency and $\beta$ the propagation constant of the mode; throughout the paper we assume and suppress a time-harmonic dependence $e^{-i\omega t}$). Conversely, fast waves ($\omega/\beta > c_0$) in open waveguides are expected to gradually lose energy in the form of leaky-wave radiation [24],[28]. In this scenario, embedded eigenstates correspond to fast waves that do not radiate even if coupling with radiation modes would be allowed (and expected) in virtue of momentum conservation.

To elucidate this situation, consider the 2D geometries sketched in Figure 1b, which are assumed to be unbounded in the *x* and *z* directions. Basic electromagnetic theory shows that, if a single 2D structure standing in a homogenous dielectric medium supports a traveling-wave electric current sheet $\mathbf{J} = \mathbf{J}_0 e^{ik_x x} \delta(y)$, with propagation constant $\left|\operatorname{Re}[k_x]\right| = |\beta| < k_0 = \omega/c_0$, then the electric current *inevitably* loses energy in the form of plane-wave radiation at an angle $\theta_{LW} = \cos^{-1}(\beta/k_0)$ from the surface [24],[28]. This implies that the transverse wavenumber is complex, i.e., $k_x = \beta - i\alpha$, where $\alpha$ is the attenuation constant accounting for energy leakage. As originally observed in [14], embedded eigenstates, i.e., non-leaky fast modes, cannot be realized in ultrathin metasurfaces composed of dense 2D arrays of polarizable elements, as they support a purely electric current distribution of this kind when externally illuminated [29]-[31]. (A relevant exception arises in the case in which all induced dipoles oscillating in phase, i.e., $\beta = 0$, and polarized normally to the array plane, which corresponds to the original example of non-radiating source distribution presented by Ehrenfest [3]). We also would like to stress that here we assume that the considered 2D and 1D structures are transversely homogenous, or, if periodic, that only one space harmonic (Floquet-Bloch mode) radiates (and no higher diffraction orders emerge when the structure is externally illuminated), so that we can evaluate the leaky-wave radiation by assuming the presence of average current sheets. Instead, for periodic structures with multiple diffraction orders, such as diffraction gratings, different anomalous effects can occur, especially at the intersection of resonant Wood's anomalies and Rayleigh anomalies, as we recently reported in [30], or near stopbands within the radiation continuum [24],[28].

Different from the case of a single electric-current sheet considered above, we have found that a 2D sheet of traveling-wave current can always be made ideally non-radiative if, in addition to the electric current $\mathbf{J} = \mathbf{J}_0 \, e^{ik_x x} \delta(y)$ considered above, we introduce a suitably tailored magnetic current $\mathbf{M} = \mathbf{M}_0 \, e^{ik_x x} \delta(y)$ with same transverse wavenumber $k_x$. An ideally non-radiating current distribution within the fast-wave region $|\beta| < k_0$ is then obtained if $\mathbf{J}_0 = J_0 \hat{\mathbf{y}}$, $\mathbf{M}_0 = M_0 \hat{\mathbf{z}}$ and

$$M_0/J_0 = -\eta_0 \beta/k_0, \qquad (1)$$

where $\varepsilon_0$ and $\eta_0$ are the free-space permittivity and impedance, respectively (the current sheets are assumed to be standing in free-space). Under these conditions (or their electromagnetic dual), the fields radiated by the electric and magnetic current sheets destructively interfere, resulting in a *fast* traveling-wave current distribution that does not radiate, despite the presence of radiation modes that would be allowed to independently carry energy away from the sources. An eigenmodal distribution of electric and magnetic currents of this kind can be induced, for example, in suitably tailored planar thick metasurfaces and photonic crystals (the presence of a transverse magnetic current always requires a finite thickness [29]) composed of periodically arranged polarizable elements with magnetic and electric dipole polarizabilities, such as high-index dielectric rods or spheres, as sketched in

Figure 1b. While all the implementations of 2D embedded eigenstates so far indeed involve periodic arrays (e.g., [14]-[17]), the discussion above shows that the periodicity of the waveguiding structure is not a necessary condition to realize embedded eigenstates. It is instead necessary that a structure – being either transversely homogenous or periodic – supports an eigenmode consistent with a non-radiating distribution of traveling-wave currents, as in (1). As a general rule, any implementation of non-radiating eigenmodes requires sufficient degrees of freedom to realize complete destructive interference, which is therefore quite challenging to achieve. Our proposed design in the following section offers a simple solution to this problem.

Analogous considerations also apply to 1D unbounded structures, such as linear arrays of particles, or cylindrical dielectric waveguides, as sketched in Figure 1c, which have also been studied in the context of embedded eigenstates [20]. Similar to the 2D case above, a linear structure supporting purely electric traveling-wave currents $\mathbf{J} = \mathbf{J}_0 e^{ik_z z} \delta(x) \delta(y)$ does not admit embedded eigenstates, consistent with passivity considerations. In fact, for linear arrays of polarizable particles with electric dipole polarizability $\alpha_{ee}$, a fast wave with real wavenumber can be supported only if $\operatorname{Im}\left[\alpha_{ee}^{-1}\right] > -k_0^3/6\pi\varepsilon_0$ [32]-[34], requiring the particles to have optical gain in order to compensate for radiation loss, and indicating that infinite lifetime without radiation cannot be achieved. However, as in the 2D case, a suitable combination of radiation contributions from electric and magnetic sources can determine a complete cancellation of radiation leakage, and hence a cylindrical embedded eigenstate [20]. It should also be pointed out that in these 1D and 2D geometries, ideally infinite Q factor is only realizable if the structure is infinite in at least one dimension, or bound by ideal mirrors (as done, e.g., in [18]).

The discussion above shows that general considerations of symmetry, dimensionality and passivity play a crucial role in the occurrence of bound states within the radiation continuum. An important take-home message of this discussion is that periodic structures (e.g., photonic crystal slabs) are not necessary to support embedded eigenstates in infinite 1D or 2D structures; instead, crucial is the ability to realize a structure (with either continuous or discrete translational symmetry) that supports traveling-wave currents of different kinds, as in Equation (1), such that the overall radiation is canceled. These concepts can be also related to the classical theory of resonant wave scattering (see, e.g., [35]-[36]). In qualitative terms, if a scattering system, in a given spectral range, exhibits only two close resonances (corresponding, for example, to the induced traveling-wave currents discussed here), then the electromagnetic eigenvalue problem defined by the homogenous wave equation can be simplified into a 2x2 complex matrix whose off-diagonal elements represent the coupling of the two resonances (the coupling depends on the specific physical/geometrical parameters of the system). Under certain conditions, one of the two complex eigenvalues may become real, hence turning a scattering resonance into a bound state in the continuum, as discussed for example in [37]-[38].

Building upon this new relevant insight, in the rest of the paper, we put forward a general platform, based on transversely-homogenous ENZ waveguides, that enables the realization of embedded eigenstates (and other electromagnetic singularities) in structures of any dimensionality, and we investigate its properties, potential, and practical implementation.

## 3. Extraordinary Light Trapping in ENZ-dielectric-ENZ Planar Waveguides

Following the discussion in the previous section, we aim at realizing 2D embedded eigenstates in planar transversely-invariant waveguiding structures. Let us consider first a transversely homogenous dielectric slab, illuminated by a plane wave impinging at an angle $\theta$. The slab exhibits periodic Fabry-Perot resonances in its reflection spectrum (**Figure 2a**), which arise when

$$k_0 d\sqrt{\varepsilon - \sin^2(\theta)} = \pi n, \qquad (2)$$

where $\varepsilon$ is the relative permittivity of the slab, $d$ its thickness, and $n = 1, 2, \ldots$ is the resonance order. Such resonances correspond to leaky modes supported by the open waveguide overdamped by radiation loss. To enhance the confinement, and therefore the radiative lifetime, of these modes, it is common to cover the slab with partially reflective screens, as typically done in leaky-wave antennas [24],[28]. In order to make a fast wave ideally confined, a simple option is to close the waveguide with perfect conductors or photonic band-gap media, hence directly forbidding the coupling with any outgoing wave. As a drastically different strategy to achieve confinement, here we take inspiration from the 3D embedded eigenstate geometries explored in [21]-[22], and we study the case in which thin plasmonic layers are symmetrically placed at the sides of a dielectric slab, as in the inset of **Figure 2b**, assuming that the permittivity of the plasmonic material follows a classic Drude dispersion $\varepsilon_{ENZ}/\varepsilon_0 = 1 - \omega_p^2/(\omega^2 + i\gamma\omega)$, where $\varepsilon_0$ is the free-space permittivity, and $\omega_p$ and $\gamma$ are, respectively, the plasma frequency and collision frequency of the electron gas. This three-layered waveguide is electromagnetically open, as the thin plasmonic layers do not prevent the fields to penetrate in the waveguide under a generic external illumination. However, if the plasma frequency is tuned near a Fabry-Perot resonance of the slab, the effect of the coatings with near-zero permittivity significantly changes the reflection spectrum of the structure, as seen in Figure 2b for transverse-magnetic (TM)-polarized plane-wave illumination (magnetic field parallel to the waveguide interfaces). We observe a sharp asymmetric variation of the reflection intensity, which arises due the interaction of the narrow-bandwidth volume plasmon resonance in the coatings and the broad Fabry-Perot resonance of the dielectric slab, similar to conventional Fano scattering resonances [39]-[40]. The sharpness of this resonant feature is a clear signature that the TM-polarized leaky mode associated with the Fabry-Perot resonance has become much more confined due to the epsilon-near-zero (ENZ) layers, which exhibit a high impedance for TM waves (the same effect can be obtained for TE waves using materials with near-zero permeability). In the lossless limit $\gamma = 0$, the Q factor of this resonant feature can actually diverge when the Fabry-Perot resonance coincides, for a certain angle, with the plasma frequency of the coatings, which results in an ideally bound mode within the radiation continuum (as we show in the following, this lossless behavior can be approximated considering low-loss zero-index metamaterial implementations). In particular, this embedded eigenstate approximately occurs at the angle of incidence

$$\theta = \sin^{-1}\left(\sqrt{\varepsilon - \left(\frac{\pi c_0 n}{\omega_p d}\right)^2}\right), \qquad (3)$$

which is therefore tunable by changing the plasma frequency $\omega_p$ of the coatings or the electrical thickness $d\sqrt{\varepsilon}$ of the dielectric slab (we assumed $\gamma = 0$). The thickness $t$ of the ENZ layers, instead, affects only marginally the position of the leaky-mode resonance in Figure 2b, while it has an effect on the dispersion of the response around the embedded eigenstate. We emphasize that the diverging resonance is due to the excitation of a bulk plasmon in the zero-index metallic layers, and not on coupled surface plasmons on the metallo-dielectric interfaces, which would be more sensitive to the layer thickness. The specific parameters of the structure considered in this example are given in the caption of Figure 2.

To better understand this light-trapping phenomenon, we have studied the dispersion $\omega(k_x)$ of the guided modes supported by the ENZ-dielectric-ENZ waveguide, applying the transverse resonance technique [24]. **Figure 2c** shows the dispersion of the propagation constant $\beta = \text{Re}[k_x]$ for the two eigenmodes of interest, which are responsible for the sharp resonant feature in the reflection spectrum. In particular, by comparing Figure 2b and 2c, we note that the scattering resonance is determined by two dispersion branches with opposite slope, corresponding to forward-propagating and backward-propagating leaky modes. The presence of an embedded eigenstate is confirmed by studying the attenuation/leakage constant $\alpha = -\text{Im}[k_x]$ of the forward-propagating branch (**Figure 2d**), which indeed vanishes at the frequency where $\varepsilon_{ENZ} \to 0$ (point I). In other words, the trajectory of a leaky pole on the complex wavenumber plane can "touch" the real axis as the frequency is varied, which corresponds to the disappearance of radiation leakage. As a result, at $\omega = \omega_p$ a pole of the system is "embedded" along the real frequency axis *within the radiation continuum*, i.e., for $|\beta| < k_0$ (fast-wave region), which represents a remarkable and unusual feature for open waveguiding systems. To respect passivity, when the structure is externally illuminated a scattering zero moves on the real axis, such that it exactly cancels the leaky pole when the latter becomes purely real, hence preventing the scattering to blow up and violate energy conservation (interestingly, we note that a similar convergence of a scattering zero and a scattering pole on the real axis may occur in active non-Hermitian systems, such as in parity-time (PT) symmetric optical systems [41], at so-called laser-absorber points [42]). Consistent with our previous discussion, the occurrence of a non-leaky fast wave can be interpreted as the realization of a *non-radiating eigenmodal distribution of polarization current*. Under external illumination, as the induced polarization current evolves toward such a non-radiating distribution, the amount of energy trapped in the open cavity grows and, in the lossless limit, it diverges at the embedded eigenstate condition. At the same time, the strong interaction of the incident wave with the ENZ-dielectric-ENZ waveguide makes the slab almost transparent, as seen in **Figure 3a**. The field distributions at the frequency of the reflection dip are shown in **Figure 3b-c**. We note that the out-of-plane magnetic field is resonantly localized within the dielectric slab, whereas the electric field is particularly strong in the ENZ layers, and it is mainly directed in the orthogonal direction, as shown by the white arrows in Figure 3c. As expected, the non-radiating eigenmodal fields supported by this transversely-invariant waveguiding configuration are indeed qualitatively consistent with the discussion in the previous section, and with the non-radiating planar source expressed by Equation (1) (traveling waves of oscillating out-of-plane magnetic dipoles and orthogonal electric dipoles). Indeed, the induced traveling-wave polarization current in the ENZ layers would not be sufficient to determine a non-radiating configuration if not for the magnetic resonance in the dielectric slab. In general, these results support the observation that the interaction of different resonators – in the present case, a magnetic cavity resonance in the dielectric slab and an electric volume-plasmon resonance in the coatings – is crucial to realize embedded eigenstates, and that this can be achieved in transversely homogenous structures, in contrast with all periodic implementations proposed in the literature. From these field distributions (and associated animations), we see that the guided mode supported by the open waveguide is indeed a *fast wave with vanishing radiation leakage*, a remarkable and counterintuitive effect. We expect that such an extreme field enhancement and frequency/angle selectivity may be useful for many applications, from advanced filtering functionalities, to enhanced selective absorption and strong light-matter interactions. We have also verified that, as it may be expected, these scattering resonances exhibit high sensitivity to small perturbations of the structure parameters, e.g., the refractive index of the dielectric region, a feature that may be exploited to realize enhanced bio-chemical sensing platforms. Conversely, the trapped state is robustly preserved when the background is perturbed (e.g., if the background is asymmetric above and below the structure), in drastic contrast with previously reported bound states within the radiation continuum [16],[43]. This behavior is expected, since the eigenmodal field distribution associated with our trapped state vanishes in the surrounding region. For example, in the case considered here, the main effect of adding a substrate below the structure would be the disappearance of the exact reflection zero near the sharp resonance feature in Fig. 2b, substituted by a reflection minimum due to the background reflection from the substrate.

**4. Local Density of Optical States of Embedded Eigenstates**

The anomalous light confinement and diverging Q factor enabled by embedded eigenstates may suggest that such trapped states may also produce an enhancement in the local density of optical states (optical LDOS), the quantity that determines the radiation and spontaneous emission rate of a localized source, which is crucial for several nano-optics applications [44]. Given the increasing interest in embedded eigenstates, there is a dire need to resolve the question of whether these structures can enable anomalously high enhancement of LDOS-mediated phenomena, such as spontaneous emission decay rates. For a source coupled to a cavity, the LDOS enhancement, or Purcell factor, is proportional to the cavity quality factor divided by the mode volume. In any 2D structure supporting embedded eigenstates, however, while the quality factor can indeed diverge, the mode volume is not easily defined, as the waveguiding structure is infinitely extended and it is electromagnetically open. Hence its LDOS cannot be intuitively predicted. Indeed, no LDOS calculation has been reported so far in the literature, since this quantity is very difficult to calculate, even numerically, due to the sampling problem in the proposed periodic geometries [45]. Conversely, our proposed design, being transversely homogenous, is the first example of a structure supporting 2D embedded eigenstates for which the LDOS, and other important electromagnetic parameters, can be calculated *analytically*, revealing general properties of these anomalous states within a rigorous theoretical framework. Without any approximation or numerical discretization, we directly calculated the LDOS for an electric dipole $\mathbf{p}$ coupled to the considered structure as the imaginary part of the analytical Green's function of the system [44]. The enhancement of power radiated by the dipole, with respect to the power $P_0$ radiated in a homogenous background with permittivity $\varepsilon_1$, , then reads

$$\frac{P}{P_0} = 1 + \frac{6\pi\varepsilon_1}{|\mathbf{p}|^2}\frac{1}{k_1^3}\mathrm{Im}\left[\mathbf{p}^* \cdot \omega^2 \mu_0 \overline{\mathbf{G}}_s(\mathbf{r}_0,\mathbf{r}_0) \cdot \mathbf{p}\right], \qquad (4)$$

where $\overline{\mathbf{G}}_s(\mathbf{r}_0,\mathbf{r}_0)$ is the scattered dyadic Green's function at the location $\mathbf{r}_0$ of the dipole (the formula above essentially calculates the rate of work done by the scattered field on the source). If we then assume that the dipole is oriented orthogonal to the planar interfaces, i.e., $\mathbf{p} = p\,\hat{\mathbf{y}}$, and after Fourier transforming the Green's function, Equation (4) can be written as

$$\frac{P}{P_0} = 1 + \frac{3}{2}\frac{1}{k_1^3}\mathrm{Re}\left[\int_0^\infty \frac{k_\rho^3}{k_{y1}} F(y_0)\,dk_\rho\right], \qquad (5)$$

where $k_1$ and $k_{y1}$ are, respectively, the total and vertical wavenumber in the region where the source is embedded, $k_\rho$ is the transverse radial wavenumber (conserved throughout the transversely homogenous structure), and $F(y_0)$ is a scalar function that includes the contributions of all the waves scattered back to the source location $y_0$, and it can be easily calculated by imposing appropriate boundary conditions at the interfaces [46], or using transmission-line theory (e.g., for a single interface, $F(y_0) = r_{TM}e^{2ik_{y1}h}$, where $h$ is the distance of the source from the interface and $r_{TM}$ is the Fresnel reflection coefficient for TM plane-wave incidence on the interface). Thanks to this spectral representation of the Green's function, we can easily identify the contribution of different plane waves and evanescent waves to the LDOS enhancement, and assess the effect of the embedded eigenstate. **Figure 4a** shows the analytically calculated wavevector-resolved LDOS (W-LDOS) enhancement (with respect to free space), as a function of frequency, for a vertical electric dipole at the center of the ENZ-dielectric-ENZ waveguide considered above, as shown in the inset. This position and orientation of the dipole has been chosen because it ensures a good overlap with the field distribution of the trapped mode within the dielectric slab, as seen in Figure 3c. Instead, the LDOS enhancement of a dipole embedded in the ENZ regions mainly stems from the strong direct excitation of volume plasmons at the plasma frequency, overshadowing the response of the guided modes of interest.

As expected, the W-LDOS plot in Figure 4a reflects the presence of guided modes (leaky and bound), consistent with Figure 2c-d. In particular, the embedded eigenstate supported by the structure is clearly visible, as the point (denoted by the roman number I) with highest and sharpest W-LDOS within the fast-wave region ($|k_\rho| < k_0$). However, when the W-LDOS is integrated over the radial wavenumber, the contribution of this trapped mode within the radiation continuum is averaged out and does not produce a peak in the spectrum of the total Purcell factor (**Figure 4b**). This is not surprising, since, from the LDOS point of view, an embedded eigenstate with $|k_\rho| < k_0$ is not different than any other conventional bound modes with $|k_\rho| > k_0$. This is also confirmed by the fact that, as seen in Figure 2c-d, the increase of radiative Q factor of these trapped modes is not typically accompanied by a decrease of group velocity $v_g = \partial \omega / \partial k$, a quantity that is inversely proportional to the LDOS in low-loss regions. Similar considerations apply also for the other 2D embedded eigenstates reported in the literature, as in [14]-[16]. Therefore, experiments aimed at probing the LDOS of a structure, such as cathodoluminescence techniques, are not expected to reveal the presence of significant features associated with 2D embedded eigenstates in the emission spectrum (the situation may be different for 3D embedded eigenstates, as those reported in [21]-[22], since these states determine a larger degree of light confinement, and smaller mode volume, than conventional bounded 3D structures). This fact implies that 2D embedded eigenstates are not expected to be beneficial for nano-optical applications aimed at controlling the radiation/emission of localized sources, whereas they may be useful when a field intensity enhancement and angle/wavelength selectivity is sought, such as for nanolasing, enhanced nonlinearities, selective absorption, etc.

Interestingly, while embedded eigenstates do not typically stand out in the Purcell factor spectrum, a different sharp peak is clearly visible in Figure 4b, at frequencies slightly below the embedded eigenstate. This corresponds to the region of the dispersion diagram, denoted by the roman number II in Figure 2 and 4a, where the two counter-propagating eigenmodes of the structure merge, going under cut-off at lower frequencies (the "kink" and merged tails of the dispersion bands in the bottom part of the diagram in Figure 2c are accompanied by very large attenuation constant, corresponding not to radiation leakage, but rather to under-cut-off behavior [28]). We observe that, in this region, the group velocity does tend to zero as the dispersion of the two eigenmodes becomes almost flat before merging, which determines a so-called Van-Hove singularity in the Green's function dispersion (see, e.g., [47]-[49]). Consistent with the discussion above, such a reduction of group velocity corresponds to an increase of the density of states, which explains the distinctive sharp peak in the LDOS spectrum in Figure 4b. Interestingly, although the Purcell factor in this example is modest, we have found that the Van-Hove singularity can be tuned and enhanced by a large extent by changing the parameters of our structure, as further discussed in the Supplementary Material. The presence of two kinds of singularities in the response of the same structure, in both the scattering spectrum (embedded eigenstates) and the radiation spectrum (Van-Hove points), clearly highlights the richness and potential of the electromagnetic properties of the ENZ-dielectric-ENZ structure analyzed in the present paper.

**5. Cylindrical Embedded Eigenstates**

Bound states in the radiation continuum have not yet been demonstrated in transversely-homogeneous 1D linear geometries, as the one in Figure 1c (top). Here, we show that the ENZ-dielectric-ENZ platform described in the previous sections can be directly translated to cylindrical coordinates, realizing an open cylindrical geometry supporting embedded eigenstates. Similar to the planar case, we expect this approach to work for waves with magnetic field parallel to the interfaces between different layers, i.e., parallel to the cylinder axis, which is typically denoted as TE polarization for cylindrical waves. In fact, for a TE cylindrical wave with magnetic field parallel to the axis of the cylindrical reference system and propagating in the radial direction $r$, the transverse wave impedance is given by $Z_n^{TE} = -i\omega\mu u_n'(kr)/u_n(kr)$, where $u_n$ is a solution of Bessel equation and the prime symbol denotes derivative with respect to the argument. By using the properties of Bessel functions, the wave impedance can be written as

$$Z_n^{TE} = i\frac{n}{r}\sqrt{\frac{\mu}{\varepsilon}} - i\omega\mu\frac{u_{n-1}(kr)}{u_n(kr)}, \qquad (6)$$

which diverges for $\varepsilon \to 0$ (the second term remains finite for small arguments, for any choice of $u_n$), with the relevant exception of the TE cylindrical harmonic of order $n=0$, associated with the magnetic dipolar contribution (which can be shown to be mostly sensitive to the permeability of the medium).

Considering a cylindrical core-shell geometry as shown in the inset of **Figure 3d**, and using Mie theory [50], we have verified that embedded eigenstates indeed exist when the permittivity of the outer layer crosses zero, while, at the same time, the core supports a TE resonance of the cylindrical cavity, analogous to the Fabry-Perot resonances in the planar case, and to embedded eigenstates in spherical open cavities [21]-[22]. The condition for cylindrical embedded eigenstates can therefore be written as

$$J_n\left(\omega_p \sqrt{\varepsilon} a/c\right) = 0, \tag{7}$$

where $\varepsilon$ and $a$ are the relative permittivity and radius of the core cylinder, respectively, $J_n$ is the Bessel function and $n = 1, 2, ...$ (we have excluded the case $n = 0$, for which a trapped state cannot be realized with purely electric material, as discussed above). Figure 3d shows an example of scattering width, for a core-shell cylinder illuminated at normal incidence, in the vicinity of the embedded eigenstate of the dominant $TE_1$ cylindrical harmonic. The trapped state manifests itself in the scattering spectrum as a sharp Fano resonance, analogous to the planar and spherical cases. The corresponding field distributions at the resonance dip are shown in **Figure 3e-f**. We observe that, near the embedded-eigenstate condition, the magnetic field is resonantly confined and dramatically enhanced within the core, whereas the electric field assumes an almost non-radiating radial distribution in the ENZ coating, and a circulating distribution in the core. Remarkably, as seen in Figure 3e, the power flow (grey streamlines) is trapped in a sort of self-sustained optical vortex within the core-shell cylinder, which is a clear signature of the external excitation of an eigenstate of the system with almost-real eigenvalue. Particularly striking is also the fact that this extreme interaction between light and a material body occurs with little scattering, as incident wavefronts and external power flow are not much perturbed by the cylinder, as indicated by the scattering dip in Figure 3d. This fact has exciting potential toward the application of these concepts to realize low-invasive light-trapping and energy-harvesting devices.

## 6. Toward a Practical Implementation with ENZ Metamaterials

In this section, we propose a realistic metamaterial implementation of the ENZ-based design introduced above for operation in the microwave frequency range. It is well known that plasmonic materials following Drude dispersion can be mimicked at low frequencies with metamaterials composed of meshes of wires [51], arrays of parallel plate waveguides [52], or textured metallic surfaces [53]. The latter case, for example, is essentially based on the fact that the phase velocity of a TE mode in a parallel-plate metallic waveguide diverges as the frequency is lowered at the cut-off condition $d/\lambda_0 = 1/(2\sqrt{\varepsilon_c})$, while, below cutoff, the propagation constant of the mode becomes purely imaginary (i.e., the mode is evanescent), hence mimicking the behavior of a wave inside a plasmonic material above and below the frequency at which the permittivity goes to zero (here, $d$ indicates the height of the parallel-plate waveguide, and the structure is filled with a material having permittivity $\varepsilon_c$).

Considering the fact that at radio-frequencies metals are very good conductors and low-loss dielectrics are available, this metamaterial platform has been used, for example, to realize plasmonic invisibility cloaks for low-loss operation at microwaves [52]. Analogously, we can apply this idea to implement the cylindrical core-shell structure presented above in order to demonstrate *near-lossless* embedded eigenstates. In particular, to realize an ENZ coating working under TE illumination, we consider a metamaterial made of an array of parallel-plate waveguides orthogonal to the cylinder axis, as shown in **Figure 5a**, designed to be exactly at cutoff at the frequency of interest. The waveguides are embedded in a host medium, which, in this example, has the same permittivity of the core cylinder, i.e., $\varepsilon_c = \varepsilon = 10$. In addition, as depicted in Figure 5a, the waveguide plates need to be separated from the inner and outer radii of the shell by a small gap $\delta$, in order to correctly mimic the field boundary condition on the interfaces, as extensively discussed in [52]. If the cut-off frequency of the waveguides is then tuned to a TE resonance of the core cylinder, according to Equation (7), and after some minor optimization, the overall structure is expected to behave as the ENZ-coated cylinder in Figure 4d-f, thereby supporting an embedded eigenstate. We have verified this fact by means of realistic numerical experiments with the commercial software CST Microwave Studio, demonstrating that the embedded eigenstate indeed manifests itself as a huge field enhancement within the structure (**Figure 5b-c**). In particular, the calculated magnetic field distribution is very similar to the ideal case in Figure 4e-f, and we observed a maximum enhancement of magnetic stored energy, inside the core cylinder, of more than 40 dB and 30 dB with respect to propagation in free-space and within the bare dielectric cylinder, respectively (Figure 5b). We expect that even higher field enhancements may be achieved if the geometry of the structure is further optimized. We would like to stress that such externally excited field intensities are quite remarkable, particularly for a structure made uniquely of non-magnetic dielectrics and metals at microwaves.

The same metamaterial platform can be considered for the implementation of the planar ENZ-dielectric-ENZ waveguides studied in the previous section. In this context, it should also be mentioned that any metamaterial implementation of this kind would unavoidably produce rapid spatial field oscillations on the surface of the structure, clearly visible in Figure 5c, which are not present in the corresponding homogenous material. This raises the question on whether this different field distribution affects the radiation properties of the metamaterial structure, compared to the ideal case. If these spatial oscillations are sufficiently fast and linearly distributed, as in infinite cylindrical and planar structures, they only produce evanescent fields that do not contribute to radiation. Instead, if the oscillations were azimuthally distributed, which would be the case in a 3D bounded structure, they would inevitably produce some radiation, consistent with the fact that any non-zero field outside a finite volume of currents corresponds to non-zero radiation [10],[21]-[23]. As a result, no metamaterial implementation of this type may be able to realize *ideal* embedded eigenstates in 3D bounded structures, and only in the limit of truly homogenous materials with vanishing or diverging constitutive parameters it may be possible to achieve a truly bounded state.

Nevertheless, from a practical standpoint, a sufficiently dense metamaterial may be designed to produce sufficiently rapid field oscillations on the surface of the structure, which would yield negligible radiation loss. Interestingly, as in the 1D and 2D scenarios described above an infinite dimension is required to ideally confine the fields and achieve a diverging Q-factor, in 3D geometries an infinitesimal granularity of the metamaterial shell is required to achieve the same effect, namely, avoid energy spilling that inevitably leads to radiation loss.

**7. Conclusion**

In this paper, we have discussed the concept of embedded eigenstates supported by different structures, shedding new light on the origin and nature of these anomalous trapped states within the radiation continuum, and highlighting their connection with non-radiating sources. Building on the physical insights we gained from our analysis of bound states in different geometries, we have devised a general platform, based on the peculiar waveguiding properties of ENZ-dielectric structures, which allows realizing embedded eigenstates of any dimensionality, and we have proposed a possible realistic implementation of cylindrical trapped states at microwave frequencies. The practical fabrication of this structure will be the focus of a future work. Our proposed design allows testing fundamental physical questions, potential, and limitations of realistic embedded eigenstate. As a relevant example, we have calculated for the first time the Purcell enhancement associated with a 2D embedded eigenstate, demonstrating no significant difference compared to a conventional bound mode. This finding helps to more clearly define which applications may benefit from embedded eigenstates.

The presence of electromagnetic singularities in the scattering and radiation spectra of the proposed structures represents a striking example of the enticing wave physics of open (non-Hermitian) systems. We note intriguing connections between structures supporting embedded eigenstates and other non-Hermitian systems such as PT-symmetric optical structures [41], as they both possess real eigenvalues, although their nature of open systems would typically imply a complex eigenspectrum, namely, damped eigenmodes due to the coupling with the available loss channels: radiation and material absorption, respectively. However, while PT-symmetric systems exhibit a continuous spectrum of real eigenvalues until a phase transition occurs (spontaneous PT-symmetry breaking), embedded eigenstates form a discrete spectrum overlapped to the continuum of radiation modes.

The extreme electromagnetic response enabled by embedded eigenstates may find application in many different scenarios, which may benefit from the spectacularly high field intensity and energy concentration inside the proposed structures (a field enhancement that is often not accompanied by enhanced scattering, ideal for low-invasive and stealth applications). For example, we speculate that by suitably loading the structure in Figure 5 with a rectifying sensor/receiver placed inside the core, the loaded structure could work as a wireless energy-harvesting device with large output power. Furthermore, the recent discovery of large optical nonlinearities in doped semiconductors (such as indium tin oxide) in their ENZ operation [54], combined with the concept of embedded eigenstate discussed here, may open interesting avenues of research, including the possibility of further boosting the nonlinear response, or using such intensity-dependent material properties to make a nonreciprocal light trapping device as proposed in [55].

To conclude, we believe that our findings shed new light on the fascinating phenomenon of embedded photonic eigenstates in open structures, and represent an enabling advance toward putting these ideas to fruition for diverse practical applications.

**Supporting Information**

Additional supporting information may be found in the online version of this article at the publisher's website.

**Acknowledgements** This work was supported by the Welch Foundation with grant No. F-1802, the Simons Foundation and the Air Force Office of Scientific Research. A.A. was partially supported by a Royal Dutch Academy of Sciences (KNAW) fellowship.

Received:
Revised:
Published online:

**Keywords**: embedded eigenstates, metamaterials, epsilon-near-zero media, scattering, leaky waves

[11] V. A. Fedotov, A. V. Rogacheva, V. Savinov, D. P. Tsai, N. I. Zheludev, Resonant transparency and non-trivial non-radiating excitations in toroidal metamaterials, *Sci. Rep.* **3,** 2967 (2013).

[12] I. Liberal, N. Engheta, Nonradiating and radiating modes excited by quantum emitters in open epsilon-near-zero cavities. *Sci. Adv.* **2**, e1600987 (2016)

[13] S. Venakides, S. P. Shipman, Resonance and Bound States in Photonic Crystal Slabs, *SIAM J. Appl. Math.* **64,** 322–342 (2003).

[14] D. Marinica, A. Borisov, S. Shabanov, Bound States in the Continuum in Photonics, *Phys. Rev. Lett.* **100,** 183902 (2008).

[15] C. W. Hsu, B. Zhen, S.-L. Chua, S. G. Johnson, J. D. Joannopoulos, M. Soljačić, Bloch surface eigenstates within the radiation continuum. *Light Sci. Appl.* **2,** e84 (2013).

[16] C. W. Hsu, B. Zhen, J. Lee, S.-L. Chua, S. G. Johnson, J. D. Joannopoulos, M. Soljačić, Observation of trapped light within the radiation continuum, *Nature* **499,** 188–91 (2013).

[17] A. Kodigala, T. Lepetit, Q. Gu, B. Bahari, Y. Fainman, B. Kanté, Lasing action from photonic bound states in continuum, *Nature* **541,** 196–199 (2017).

[18] T. Lepetit, B. Kanté, Controlling multipolar radiation with symmetries for electromagnetic bound states in the continuum, *Phys. Rev. B* **90,** 241103 (2014).

[19] E. N. Bulgakov, A. F. Sadreev, A. F. Bloch bound states in the radiation continuum in a periodic array of dielectric rods, *Phys. Rev. A* **90,** 053801 (2014).

[20] E. N. Bulgakov, A. F. Sadreev, Light trapping above the light cone in a one-dimensional array of dielectric spheres, *Phys. Rev. A* **92,** 023816 (2015).

[21] M. G. Silveirinha, Trapping light in open plasmonic nanostructures, *Phys. Rev. A* **89,** 023813 (2014).

[22] F. Monticone, A. Alù, Embedded Photonic Eigenvalues in 3D Nanostructures, *Phys. Rev. Lett.* **112,** 213903 (2014).

[23] A. J. Devaney, E. Wolf, Radiating and Nonradiating Classical Current Distributions and the Fields They Generate. *Phys. Rev. D* **8,** 1044–1047 (1973).

[24] F. Monticone, A. Alu, Leaky-Wave Theory, Techniques, and Applications: From Microwaves to Visible Frequencies. *Proc. IEEE* **103,** 793–821 (2015).
24

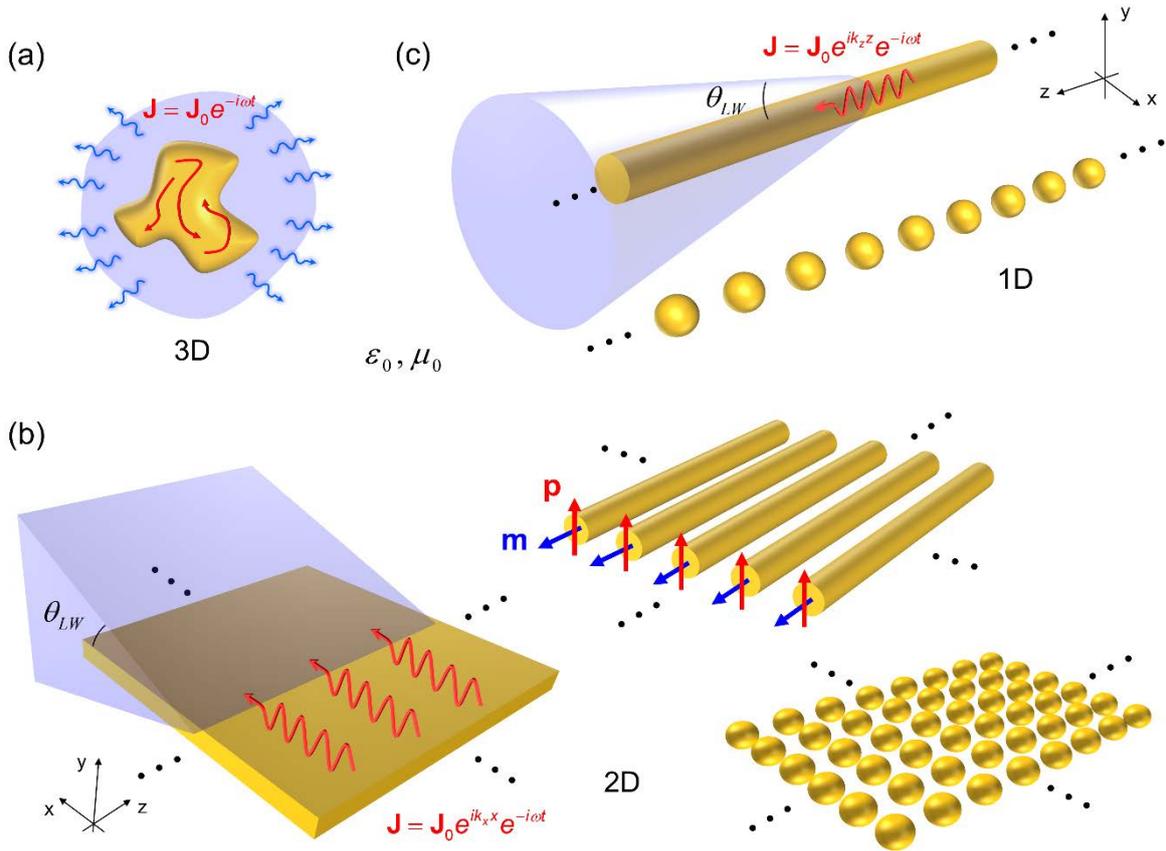

**Figure 1. Open cavities of different dimensionality and symmetry.** (a) 3D bounded open cavity supporting an oscillating current **J** gradually losing energy in the form of spherical-wave radiation (denoted by the blue area and arrows) in the background medium. (b) 2D open wave-guiding structures with continuous translational symmetry (left) and discrete

translational symmetry in one (center) and two directions (right). The structure on the left is shown to support a traveling wave current decaying in time and space, the energy being lost in the form of planar leaky-wave radiation, represented by the blue shape. The structure at the center supports electric **p** and magnetic **m** dipoles, with the required orientation to support a non-leaky fast wave. (c) Similar to (b) but for a 1D open wave-guiding structure, with continuous (top) and discrete (bottom) translational symmetry in the *z* direction. The dots in (b) and (c) indicate the directions for which a structure is infinitely extended. All structures are assumed to be surrounded by free-space.

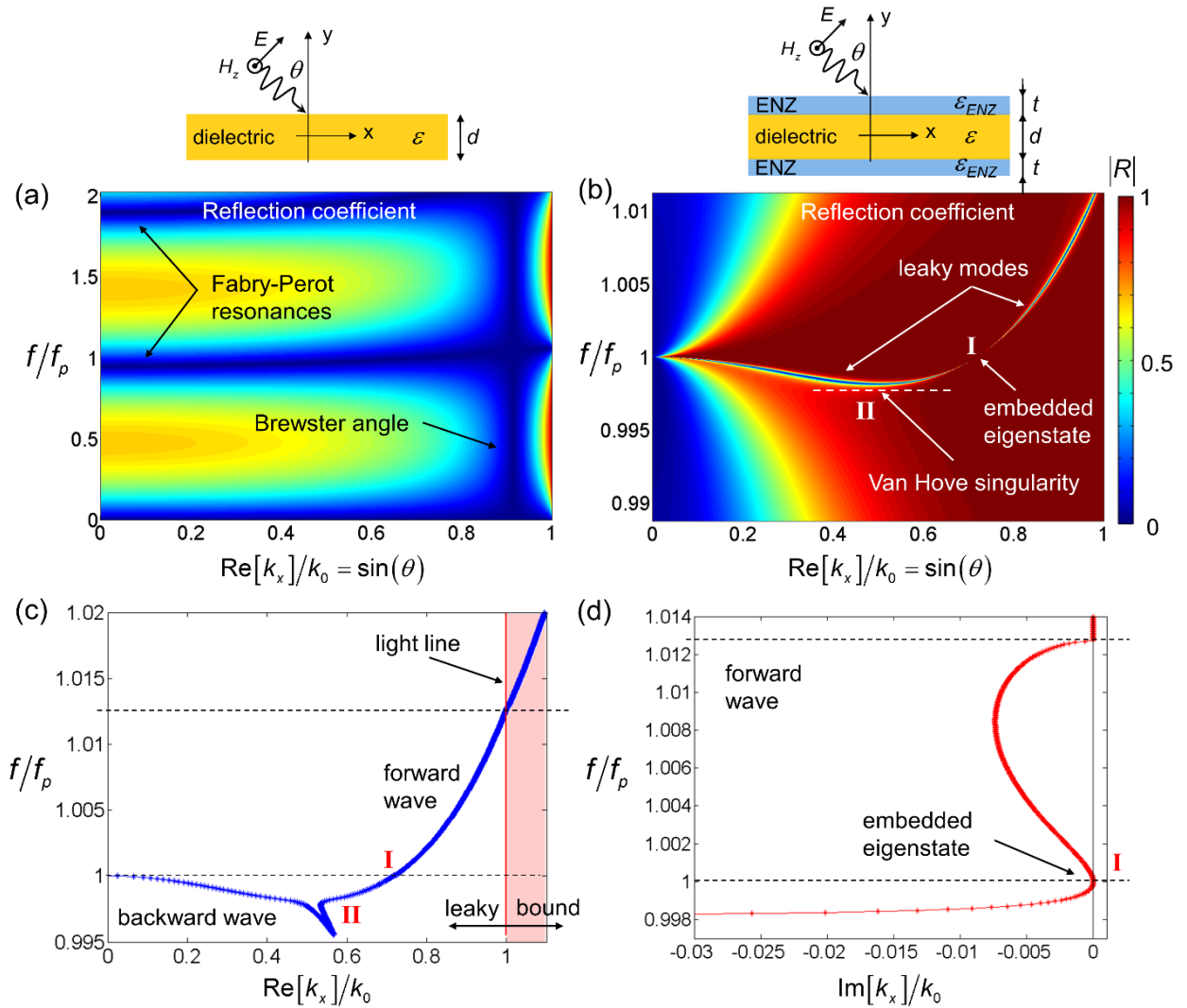

**Figure 2. Embedded Eigenstate in a Planar ENZ-dielectric-ENZ Waveguide.** (a-b) Amplitude of the reflection coefficient for a transversely homogenous dielectric slab of permittivity $\varepsilon = 5$ and thickness $d = 1$ $\mu$m, as a function of frequency and transverse wavenumber, under TM-polarized plane-wave illumination, as depicted in the inset. The frequency is normalized to the plasma frequency of the ENZ layers considered in panel (b), $f_p = 70.8$ THz. (b) Similar to (a) but for the same dielectric slab covered by ENZ layers, as shown in the inset, with thickness $t = 0.1$ $\mu$m. (c) Frequency dispersion of the real part of the wavenumber for two eigenmodes of the structure in panel (b). The vertical red line represents the light line, which divides fast- and slow-wave regions. The horizontal dashed lines indicate the plasma frequency and the frequency at which the forward-propagating eigenmode enters the slow-wave region. (d) Similar to (c), but for the imaginary part of the wavenumber of the forward eigenmode. Roman numerals I and II in panels (c) and (d) indicate two regions of interest of the dispersion diagram, around an embedded eigenstate (I), and where the two dispersion branches become flat before merging (II), as discussed in the text.

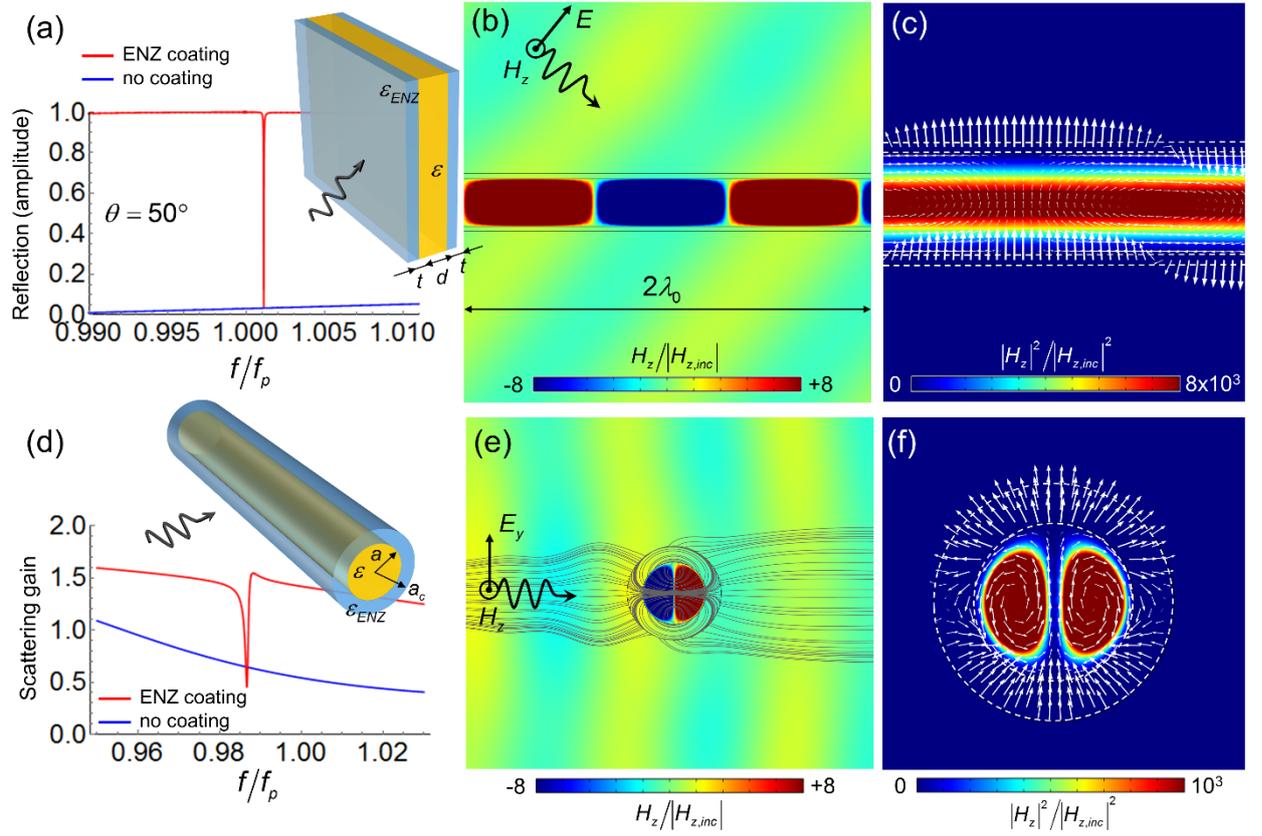

**Figure 3. Field Distributions of Embedded Eigenstates.** (a) Spectrum of the reflection coefficient (amplitude) for the ENZ-dielectric-ENZ waveguide considered in Figure 2 (red line, sketch in the inset), and of the bare dielectric slab (blue line), under TM-polarized plane-wave illumination at an angle of $\theta = 50°$ with respect to the surface normal. (b) Corresponding time-snapshot of the magnetic field distribution, and (c) magnetic field intensity (colors) and time-snapshot of the electric field distribution (white arrows), all at the frequency of the reflection minimum in (a). The structure in (a-c) is infinitely extended in the lateral directions. (d) Spectrum of the scattering width (normalized to the cylinder outer diameter) for the core-shell cylinder shown in the inset (red line), and for the bare core cylinder (blue line), with permittivity $\varepsilon = 10$, aspect ratio $a_c/a = 1.5$ and $a = 0.197\lambda_p$, where $\lambda_p$ is the free-space wavelength at the plasma frequency $f_p$ of the ENZ coating. The cylinders are normally illuminated by a TE-polarized plane wave (magnetic field parallel to the cylinder axis). (e) Corresponding time-snapshot of the magnetic field distribution, and (f) magnetic field intensity (colors) and time-snapshot of the electric field distribution (white arrows), all at the frequency of the scattering dip in (d). Gray lines in (e) represents the power density flow (time-averaged Poynting vector) near the cylinder.

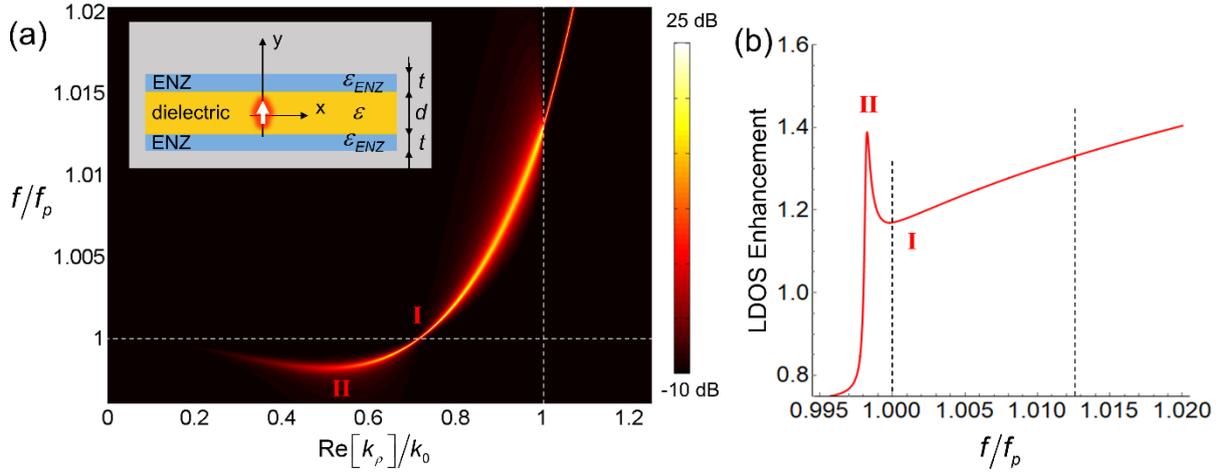

**Figure 4. Local Density of Optical States in Planar ENZ-dielectric-ENZ waveguides.** (a) Wavevector-resolved LDOS enhancement (with respect to vacuum) of the structure considered in Figure 2, as a function of frequency normalized to the plasma frequency $f_p$ of the ENZ layers, for the case of a vertical electric dipole at the center of the dielectric slab, as shown in the inset. The ENZ layers are slightly lossy, with collision frequency $\gamma = 10^{-5}\omega_p$. All the other parameters are the same as in Figure 2. The vertical dashed line indicates the border between fast-wave and slow-wave regions. (b) Corresponding LDOS enhancement with respect to vacuum. The two vertical dashed lines indicate the plasma frequency and the frequency at which the mode enters the slow-wave region.

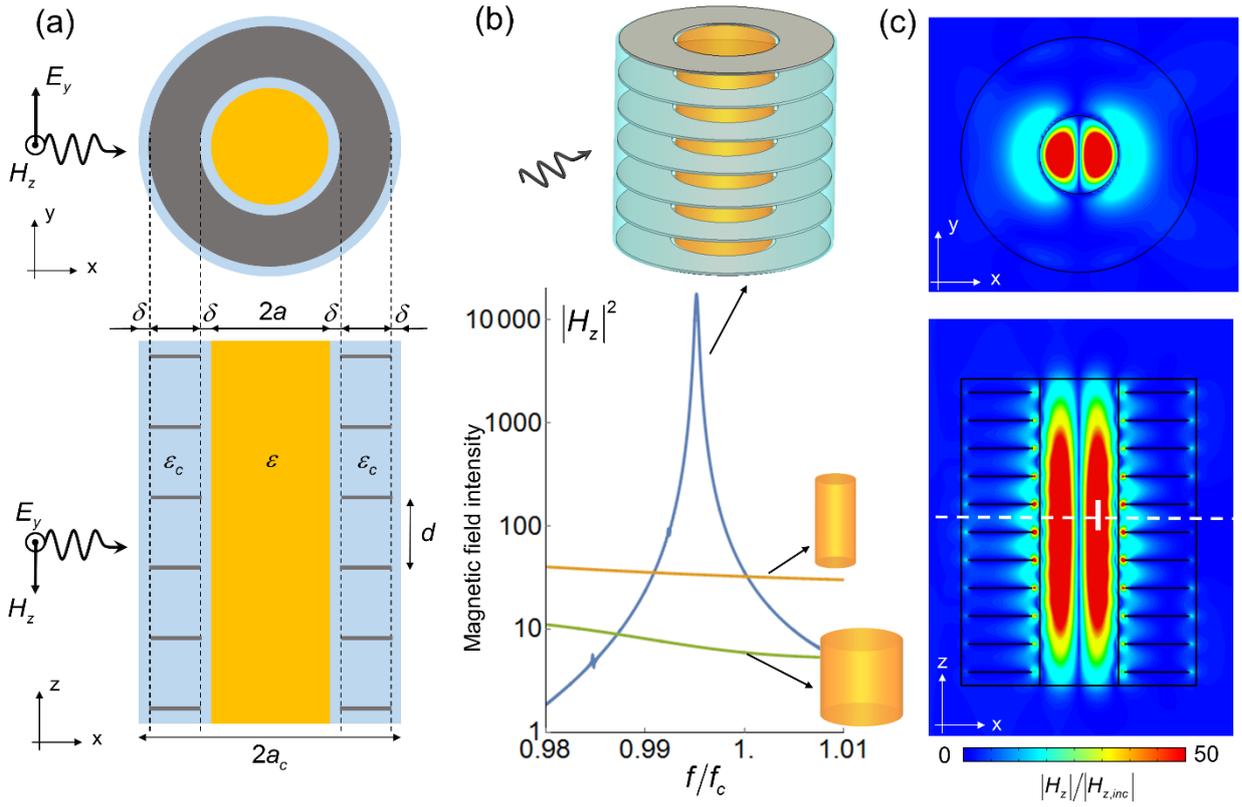

**Figure 5. Implementation of Cylindrical Embedded Eigenstates based on ENZ Metamaterials.** (a) Sections in the xy plane (top) and the xz plane (bottom) of the considered cylindrical structure illuminated by a TE polarized plane wave, as indicated on the left. A 3D view of the structure is shown in the top inset of panel (b). The parameters are: $a = 5.9$ cm, $a_c = 3a$, $d = 4.23$ cm, $\delta = 0.25d$, and $\varepsilon = \varepsilon_c = 10$. Grey regions are made of metal, assumed perfectly conducting. The central frequency is $f_c = 1$ GHz, at which the metallic waveguides are at cutoff. (b) Maximum magnetic field intensity enhancement (blue line), with respect to free-space, inside the core-shell cylinder in (a), at the position indicated by the vertical white segment in (c). For comparison, the plot shows also the maximum magnetic field intensity enhancement for the same structure without metallic plates (green line), and for the bare core cylinder (orange). (c) Distribution of the magnetic field intensity inside the structure in (a), at the frequency corresponding to the maximum in (b). The fields are plotted on the xz plane (bottom) and the xy plane (top; dashed white line in the xz plot).

**Graphical Abstract**

We discuss the possibility of trapping light in open structures based on the concept of bound states within the radiation continuum enabled by zero-index metamaterials. Relevant properties of these anomalous trapped states are discussed in different settings, and a general platform is put forward that allows realizing extremely-high field enhancements in open structures under external illumination, paving the way for the application of these concepts in practical scenarios.

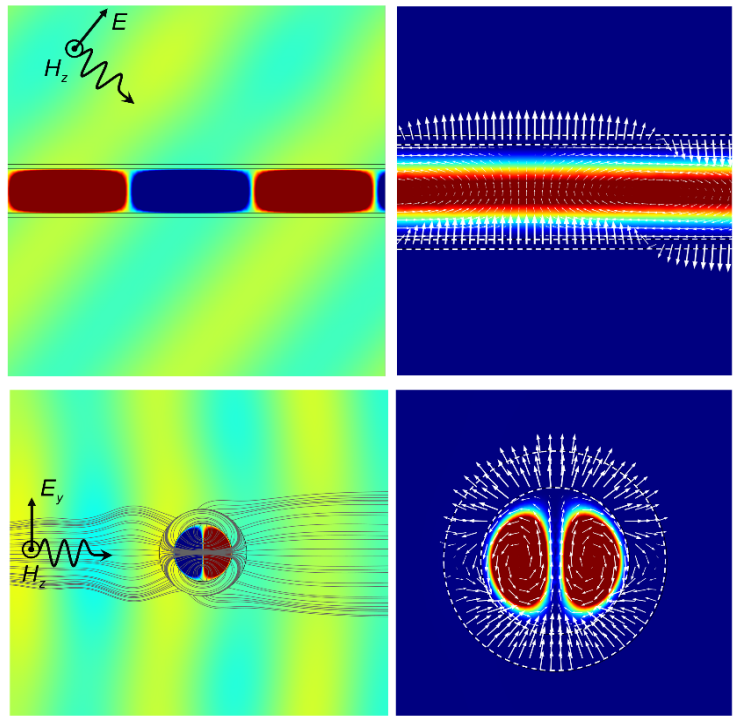